\documentclass[sigconf]{acmart}
\usepackage{balance}
\usepackage{booktabs} % For formal tables
\usepackage[german,american]{babel}
\usepackage[T1]{fontenc}
\usepackage[latin1]{inputenc} % For umlauts (OT1) and the ß (OT1, T1).
\usepackage{array}
\usepackage{hyperref}
\usepackage{xcolor}

\newcommand{\Ni}{(1)~}
\newcommand{\Nii}{(2)~}
\newcommand{\Niii}{(3)~}
\newcommand{\Niv}{(4)~}
\newcommand{\Nv}{(5)~}
\newcommand{\Na}{(a)~}
\newcommand{\Nb}{(b)~}

\RequirePackage{color}
\RequirePackage{soul}
\setstcolor{blue}

\usepackage{lscape, float} %allows vertical tables
\usepackage{graphicx}
\usepackage{ifpdf}
\ifpdf
\DeclareGraphicsExtensions{.pdf,.ai,.jpg,.png}
\else
\DeclareGraphicsExtensions{.ai,.ps,.eps}
\fi
\setkeys{Gin}{pagebox=artbox}% Alternative for \pdfpagebox 5 in preceding line.
\graphicspath{{./ictir21-axiomatic-explanations-figures/}}
\usepackage{xspace}

\newcommand{\tf}{\ensuremath{\mathit{tf}}}

\begin{document}
\settopmatter{printfolios=true}
\title{Towards Axiomatic Explanations for Neural Ranking Models}

\def\bsauthor{%
\mbox{\hspace*{-2.1em}
Michael V{\"o}lske,$^{*}$
Alexander Bondarenko,$^{\dagger}$
Maik Fr{\"o}be,$^{\dagger}$
Benno Stein,$^{*}$
\hspace*{-2.1em}}\\
\mbox{\hspace*{-2.1em}
Jaspreet Singh,$^{\ddagger}$
Matthias Hagen,$^{\dagger}$
Avishek Anand$^{\S}$
\hspace*{-2.1em}}

\gdef\bsauthor{%
Michael V{\"o}lske,
Alexander Bondarenko,
Maik Fr{\"o}be,
Benno Stein,
Jaspreet Singh,
Matthias Hagen,
Avishek Anand}}

\author[
M.~V{\"o}lske,
A.~Bondarenko,
M.~Fr{\"o}be,
B.~Stein,
J.~Singh,
M.~Hagen,
A.~Anand
]{\bsauthor}

\affiliation{%
%\fontsize{9}{10}\selectfont
\mbox{\kern-0.5em$^{*}$}Bauhaus-Universit{\"a}t Weimar\quad%\quad\;
\mbox{\kern-0.5em$^{\dagger}$}Martin-Luther-Universit{\"a}t Halle-Wittenberg\quad%\quad\;
\mbox{\kern-0.5em$^{\ddagger}$}Amazon\quad%\quad\;
\mbox{\kern-0.5em$^{\S}$}Leibniz Universit{\"a}t Hannover 
\institution{}
\city{}
\country{}
}

\begin{abstract}

Recently, neural networks have been successfully employed to improve upon state-of-the-art effectiveness in ad-hoc retrieval tasks via machine-learned ranking functions. While neural retrieval models grow in complexity and impact, little is understood about their correspondence with well-studied IR principles. Recent work on interpretability in machine learning has provided tools and techniques to understand neural models in general, yet there has been little progress towards explaining ranking models.

We investigate whether one can explain the behavior of neural ranking models in terms of their congruence with well understood principles of document ranking by using established theories from axiomatic~IR. Axiomatic analysis of information retrieval models has formalized a set of constraints on ranking decisions that reasonable retrieval models should fulfill. We operationalize this axiomatic thinking to reproduce rankings based on combinations of elementary constraints. This allows us to investigate to what extent the ranking decisions of neural rankers can be explained in terms of the existing retrieval axioms, and which axioms apply in which situations. Our experimental study considers a comprehensive set of axioms over several representative neural rankers. While the existing axioms can already explain the particularly confident ranking decisions rather well, future work should extend the axiom set to also cover the other still ``unexplainable'' neural~IR rank decisions.
\end{abstract}




\copyrightyear{2021}
\acmYear{2021}
\setcopyright{acmlicensed}\acmConference[ICTIR '21]{Proceedings of the 2021 ACM
SIGIR International Conference on the Theory of Information Retrieval}{July 11,
2021}{Virtual Event, Canada}
\acmBooktitle{Proceedings of the 2021 ACM SIGIR International Conference on the
Theory of Information Retrieval (ICTIR '21), July 11, 2021, Virtual Event,
Canada}
\acmPrice{15.00}
\acmDOI{10.1145/3471158.3472256}
\acmISBN{978-1-4503-8611-1/21/07}

% demanded by https://www.scomminc.com/pp/acmsig/ictir.htm
\fancyhead{}
\settopmatter{printacmref=true}

\maketitle
\section{Introduction}
\label{sec:intro}

When using machine learning models to rank search results, the training data (clicks, human annotations) drives the way features are combined as relevance signals by various models, ranging from linear regression and decision trees to deep neural networks more recently~\cite{mitra:2017,nogueira:2020,rudra2020distant}. On a more abstract level, by learning how to combine features to best rank documents, a machine-learned model indirectly encodes the query intent. Documents are then ordered by relevance, i.e., how well they match the underlying query intent. While such models have achieved state-of-the-art effectiveness in ad-hoc text retrieval, their complexity makes them difficult to interpret, up to the point that eventually some of the reported performance gains have been called into question~\cite{lipton:2019}.

Axiomatic thinking on relevance scoring functions has arisen from a similar concern regarding a lack of rigor in formulating what makes a good result ranking. Empirically, non-optimal parameter settings had been shown to cause existing retrieval models to perform poorly, resulting in heavy parameter tuning. Axiomatic practitioners have hence formalized desirable, elementary properties of ranking functions.
The analysis of popular scoring functions with respect to their adherence to axiomatic constraints has given rise to revised scoring functions with provably superior performance~\cite{lv:2011}, re-ranking approaches that improve an existing ranking's adherence~\cite{stein:2016n}, as well as datasets for diagnosing retrieval models' axiom adherence empirically~\cite{rennings:2019,camara:2020}.

In this paper, we apply established retrieval axioms to ground the behavior of arbitrary ranking models in a well understood (and hence interpretable) axiomatic basis. The central question that we raise in this paper is: ``To what extent can we explain neural models in terms of the existing IR~axioms?'' To this end, we follow Hagen~et~al.~\cite{stein:2016n} and operationalize the retrieval axioms as Boolean predicates that, given a pair of documents and a query, express a preference for either one document or the other to be ranked higher. Given multiple axioms, this yields a set of possibly conflicting ranking preferences for each document pair. Thus, to explain a given result ranking, we compute axiom preferences across all of its constituent document pairs and then fit an explanation model, which is here a simple classification model trained to make the same pairwise ordering decisions as the initial ranking, given only the axiom preferences as predictors. This approach permits various insights into the model that produced the initial ranking: \Na the explanation model's parameters reveal the degree to which different axiomatic constraints are important to the retrieval model under consideration, and \Nb the fidelity with which the initial ranking can be reconstructed can point to blind spots in the axiom set, which can help to uncover new ranking properties yet to be formalized.

Several previous studies have used retrieval axioms to explain or improve ranking decisions, but our work is the first to combine many axioms together to specifically try to reconstruct neural rankings in an extensive experimental study. Our paper makes the following contributions:~%
\Ni We propose a general-purpose framework to analyze arbitrary rankings with respect to their adherence to information retrieval axioms. \Nii In an extensive experimental study on the Robust04~\cite{voorhees:2004} and MS~MARCO~\cite{nguyen:2016} test collections, we investigate to what extent five different state-of-the-art neural retrieval models can be explained under the axiomatic framework.\footnote{%
	For the sake of reproducibility, the code for the experiments described in this paper is made available at \url{https://github.com/webis-de/ICTIR-21}
}
\Niii We explore notions of locality in axiomatic explanations, such as whether different explanations apply to different queries, or to different locations in result rankings.

Our results show that the degree to which rankings can be explained with the currently known axioms is still rather limited overall. The axiomatic explainability of neural rankers tends to be on par with that of simpler classical retrieval models. Large differences in the retrieval score, where retrieval functions are highly confident of the relevance differences they indicate, are well explainable with axioms, but this is much less the case for small fluctuations among closer result documents. Interesting directions for future work thus are the formulation of additional axioms that capture other angles of relevance and that apply in a wide range of real-world contexts, as well as the development of a more rigorous approach to relaxing the preconditions of the existing axioms for a wider applicability.%
\section{Related Work}
\label{sec:related-work}

The fundamental task of information retrieval---extracting from a large collection those information items relevant to a particular information need---is typically accomplished by ranking the items in a collection according to assigned relevance scores. Most commonly-used scoring functions such as~BM25 have been designed to quantify some specific, narrow notion of relevance~\cite{fang:2005,amigo:2017}. Unlike recently-popular machine learning models, axiomatic IR aims to explain a ``good'' retrieval function by means of mathematically described formal constraints. Possibly the earliest ideas that are close to ``axiomatic'' IR were a retrieval system complemented with the production rules from artificial intelligence~\cite{mccune:1985}, which improved the performance of a Boolean model, a formalization of a conditional logic underlying information retrieval~\cite{rijsbergen:1986}, and terminological logic to model IR processes~\cite{meghini:1993}. The first real mention of the term axiom in relation to IR was introduced in a study by \citeauthor{bruza:1994}~\cite{bruza:1994}, who proposed to describe retrieval mechanisms by axioms expressed in terms of concepts from the information field.

In the last decades, the number of studies developing new axioms that describe what a good retrieval function looks like, has considerably increased. More than 20 distinct axioms have been proposed so far, which can be divided into groups by  the particular aspect of the relevance scoring problem that they aim to formalize: term frequency~\cite{fang:2004,fang:2005,fang:2011,na:2008} and lower bounds on it~\cite{lv:2011,lv:2012}, document length~\cite{fang:2004,cummins:2012}, query aspects~\cite{zheng:2010,wu:2012,gollapudi:2009}, semantic similarity~\cite{fang:2006,fang:2008}, term proximity~\cite{tao:2007,stein:2016n}, axioms for evaluation~\cite{amigo:2013,busin:2013}, axioms describing properties implied by link graphs~\cite{altman:2005}, axioms for learned ranking functions~\cite{cummins:2007,cummins:2011}, multi-criteria relevance scoring~\cite{gerani:2012}, user-rating based ranking~\cite{zhang:2011}, translation language model axioms~\cite{karimzadehgan:2012,rahimi:2014}, and term dependency~\cite{ding:2008}. The majority of the aforementioned studies consider the axioms individually. 

The first large-scale study on axioms' impact on retrieval effectiveness was published by~Hagen~et~al.~\cite{stein:2016n}, who combined 23~individual axioms to re-rank top-\emph{k} result sets and showed that different axiom combinations significantly improve the retrieval performance of basic models such as~BM25, Terrier DPH, or DirichletLM.
We follow this approach of operationalizing retrieval axioms as computable functions, which enables the exploitation of a large set of axioms. However, instead of manipulating rankings to more closely match axiomatic preferences, we reconstruct a retrieval model's rankings via weighted axiom preferences to better understand and explain the decisions underlying the ranking function.

\begin{table}[tb]
\caption{The 20 retrieval axioms included in our study; STMC1 and STMC2 each implemented in three variants ($^*$).}
\label{table:axioms}
\centering
\small
{\renewcommand{\arraystretch}{1.2}
\begin{tabular}{@{}ll@{}}
\toprule
\bfseries Purpose             & \bfseries Axioms / Sources      \\
\midrule
Term frequency                & TFC1~\cite{fang:2004},     
                                TFC3~\cite{fang:2011},     
                                 TDC~\cite{fang:2004}      \\
Document length               &   LNC1~\cite{fang:2004},
                                TF-LNC~\cite{fang:2004}      \\
Lower-bounding term frequency & LB1~\cite{lv:2011}        \\
Query aspects                 & REG~\cite{wu:2012},          
                                AND~\cite{zheng:2010},       
                                DIV~\cite{gollapudi:2009} \\
Semantic Similarity           & STMC1$^*$~\cite{fang:2006},        
                                STMC2$^*$~\cite{fang:2006}      \\
Term proximity                & PROX1--PROX5~\cite{stein:2016n} \\
%Other axioms                  & ORIG~\cite{stein:2016n} \emph{(not used)}    \\

\bottomrule
\end{tabular}
}
\end{table}

Recently, retrieval axioms have been used to regularize neural retrieval models to prevent over-parameterization, improving both training time and retrieval effectiveness~\cite{rosset:2019}. Retrieval axioms have also been applied to weight meta-learner features which predict how to combine the relevance scores of different retrieval models into an overall score~\cite{arora:2019}. Rennings~et~al.~\cite{rennings:2019} have developed a pipeline to create diagnostic datasets for neural retrieval models, each fulfilling one axiom, which allows to detect what kind of axiomatically expressed search heuristics neural models are able to learn. The study's diagnostic datasets focus on only~4 simple individual axioms, which cannot completely account for neural rankers' decisions.
In a follow-up publication, C{\^{a}}mara and Hauff~\cite{camara:2020} extend the idea to building diagnostic datasets for~9 axioms separately, with a focus on BERT-based rankers.
MacAvaney et al.~\cite{macavaney:2020} systematize the analysis of neural IR models as a framework comprising three testing strategies---controlled manipulation of individual measurements (e.g., term frequency or document length), manipulating document texts, and constructing tests from non-IR datasets---whose influence on neural rankers' behavior can be investigated.
In our study, we follow a fourth approach: reconstructing rankings based on elementary axiomatic properties. In the process, we combine the ideas of 20~retrieval axioms at the same time (cf.\ Table~\ref{table:axioms}; variants of STMC1 and STMC2 are described in Section~\ref{subsec:operationalizing-retrieval-axioms}), aiming to capture and explain arbitrary neural rankers' decisions. While the aforementioned studies on axiomatic diagnostics of neural rankers typically operate in very controlled and synthetic settings, our approach provides a complementary view based on more realistic, TREC-style queries and document collections.

The rationales behind the decisions of complex learning systems are also studied in the field of interpretability in machine learning~\citep{lakkaraju:2016,zhang2021explain:expred}. 
Recent work on extracting feature attributions using post-hoc approximations is similar to our model of explaining document preference pairs~\citep{singh:2018,singh:2020,singh2019exs,fernando2019study:shap}. However, we crucially differ from them in two ways: we use axioms as possible explanations and we employ a learning framework to measure the fidelity to the original model rather than a combinatorial framework~\citep{singh:2018}.

\begin{figure*}[tb]
	\centering
	\includegraphics[width=0.9\textwidth]{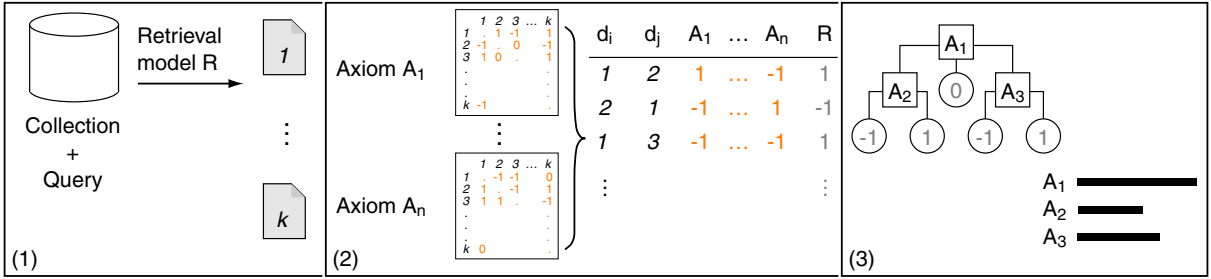}
	\caption{Overview of the axiomatic explanation pipeline. \Ni The retrieval model R ranks a document set of size~\emph{k}. \Nii Axioms produce ranking preferences for all document pairs. \Niii A simple explanation model, trained to recreate the ordering produced by R using the axiom preferences as features, reveals which axioms generate the ranking.}
	\label{fig:explanation-approach}
\end{figure*}

\section{Axiomatically Explaining Rankings}
\label{sec:approach}

Our axiomatic explanation framework generates post-hoc explanations for the ranked result lists produced by the retrieval model under investigation. The two main components are a set of \emph{axioms}, and an \emph{explanation model}. Figure~\ref{fig:explanation-approach} provides a high-level overview of the framework: Given a ranked list of documents $(d_1,\dots, d_k)$ and a set of axioms $\{A_1,\dots, A_n\}$, we first compute the pairwise ranking preferences for every document pair under each axiom.
Thus, each document pair $(d_i, d_j)$ is associated with an $n$-dimensional vector of axiom preferences, along with its ordering in the original ranking. The axioms are operationalized as predicates that map a given document pair to a ternary ranking preference---prefer $d_i$, prefer $d_j$, or prefer neither. The explanation model aggregates a vector of such axiomatic ranking preferences to a final ordering for a given document pair in such a way that the ranking decisions of the retrieval model under investigation are reproduced as faithfully as possible. The explanation model's parameters give insights into how the retrieval axioms contribute to the ranking under scrutiny.

\subsection{Operationalizing Retrieval Axioms}\label{subsec:operationalizing-retrieval-axioms}

While a variety of axioms for different aspects of retrieval---such as ranking, evaluation, or relevance feedback---have been specified in the literature, we include only those that can be restated to express ranking preferences on pairs of documents. In operationalizing those axioms, we follow the approach of Hagen~et~al.~\cite{stein:2016n}, but make modifications to adapt their axiomatic re-ranking framework to our ranking explanation setting. In our setting, each axiom~$A$ implements a ternary predicate $A(d_i, d_j, q)$ that, given a document pair~$(d_i, d_j)$ and query~$q$, maps to a ranking preference taking on values of~$1$, $-1$ or $0$ depending on whether the axiom would rank document $d_i$ higher, $d_j$ higher, or has no preference on the pair.

Table~\ref{table:axioms} summarizes the retrieval axioms we employ in our study grouped by the general notion of relevance that they capture (axioms missing from the table cannot easily be restated to express actual ranking preferences). For example, the term frequency axioms TFC1, TFC3, and TDC constrain how the term frequency~\tf\ should manifest in document ranking and the first of these, TFC1, states that, given two documents of the same length and a single-term query, the document with more occurrences of the query term should be ranked higher~\cite{fang:2004}.

The axioms are generally framed in artificial preconditions to allow for precisely reasoning about the properties of retrieval functions (e.g., TFC1 for documents with exactly the same length). However, since this limits their practical applicability, we make modifications following Hagen~et~al.~\cite{stein:2016n}. For example, in case of TFC1, \Ni we relax the equality constraint (i.e., we consider all document pairs with a length difference of at most~10\%), \Nii we strengthen the inequality constraint (i.e., requiring at least a 10\%~difference in term frequency), and \Niii we generalize it to multi-term queries (i.e., using the sum of term frequencies over all query terms). The other term frequency axioms are modified in a similar way. As originally stated, given two equally discriminative query terms and two same-length documents, TFC3 prefers a document containing both terms over another that contains only one whereas for two query terms of differing discriminativeness, TDC prefers the document containing the more discriminative term~\cite{fang:2011}. While implementing term discriminativeness by inverse document frequency, we again relax the equalities and strengthen the inequalities as for TFC1 and generalize to summing over more query term pairs.

For the axioms capturing document length, lower-bounding, and query aspect constraints, we largely follow the operationalization of Hagen~et~al.~\cite{stein:2016n}. In short, LNC1 prefers the shorter document given identical term frequency of all query terms, while TF-LNC prefers the document with more query term occurrences assuming the term frequencies of all non-query terms are the same. The lower-bounding axiom LB1 applies when there is a query term~$t$ such that both documents obtain the same retrieval score if~$t$ is removed from the query; in this case, LB1 prefers documents that contain~$t$ over those that do not~\cite{lv:2011}. The query aspect axiom~REG evaluates the pairwise semantic similarity of the query terms, and prefers documents with more occurrences of the term that is least similar to all others---in our experiments, we employ the Wu-Palmer measure for term similarity~\cite{wu:1994}---whereas the axiom AND~prefers documents that contain every query term at least once. The diversity axiom DIV~prefers the document that is less similar to the query (measured via Jaccard similarity in our implementation).

The semantic similarity axiom STMC1~favors documents containing terms that are semantically more similar to the query terms, whereas STMC2~requires that a document exactly matching a query term once contributes to the score at least as much as matching semantically related terms arbitrarily many times instead~\cite{fang:2006}. We operationalize STMC1 and STMC2 based on the Wu-Palmer measure in the same way as Hagen~et~al.~\cite{stein:2016n} but also additionally explore word em\-bed\-ding-based term similarity measures. We thus also have STMC1-f and STMC2-f that utilize 1~million fastText word vectors pre-trained with subword information on Wikipedia from 2017~\cite{mikolov:2018} and
STMC1-fr and \mbox{STMC2-fr} that utilize custom \mbox{fastText} embeddings trained on the Robust04 document collection. The term proximity axioms PROX1--PROX5 are employed in the same way as originally proposed by Hagen~et~al.~\cite{stein:2016n}. Beyond these, Hagen~et~al.~\cite{stein:2016n} also propose the axiom~ORIG that simply reproduces the ranking preferences of the original retrieval model. While this axiom does not immediately apply to the explanation setting, we do include its retrieval model-specific versions in some experiments in order to study how one retrieval model's decisions might be explained in terms of those of another.

\subsection{Aggregating Axiom Preferences}

Once the axiomatic ranking preferences have been computed for a set of axioms and document pairs, we fit an explanation model to reconstruct the original ranking based on the axiomatic preferences. On the level of pairwise ranking decisions, this is a binary classification task. While a wide range of different models can accomplish this task, there is a clear case for using models that are simple enough to be inspected for insights on how exactly the retrieval axioms interact to generate the ranking under scrutiny. There is a trade-off between the fidelity of the explanation model---i.e., how well it reconstructs input rankings---and the degree to which it can be inspected: higher-capacity models tend to be less interpretable.

The goal of our present study is to test the feasibility of the axiomatic explanation approach and to examine the completeness of the axiom set. We thus employ a random forest model for axiom preference aggregation. This is at the higher-fidelity end of the spectrum but still offers useful, though limited, facilities for inspection in terms of feature importance. Once a more comprehensive axiom set is established---the need for this is indicated by our current results---, lower-capacity models with richer interpretability become reasonable. For instance, in a linear model like logistic regression, both the magnitude and signs of the parameter vector would be meaningful, while a single shallow decision tree could be interpreted as a Boolean formula that explains a given ranking.

\section{Experimental Setup}
\label{sec:setup}

The experimental study described in this section focuses on two research questions: \Ni To what extent can the axioms currently known from axiomatic IR faithfully reconstruct the decisions made by neural ranking models, and how does this compare to classical retrieval scoring functions? \Nii Which retrieval axioms are most important in what scenarios, and what is the relationship between axiomatic explanations and ranking quality?

In order to answer these questions, we apply our axiomatic explanation framework on two standard evaluation datasets, and examine the top~1000 retrieved results across~200 different queries with three classical and five neural machine-learned retrieval models.
For each ranking, we sample document pairs from the set of all pairwise orderings, and generate axiomatic ranking preferences. On these, we fit a random forest-based explanation model.

\subsection{Collections}
\label{sec:collections}
\label{sec:robust04}
\label{sec:ms-marco}

Neural ranking models need large amounts of labeled training data to produce rankings substantially superior to classical retrieval models~\cite{yang:2019}, but not many training collections are available. We follow recent trends and use the Robust04 and MS~MARCO datasets. Both are standard TREC collections containing relevance judgments and due to their distinctive size and characteristics---in terms of constituent documents and the associated search queries---allow us to test our approach in various scenarios.

The Robust04 dataset was developed to improve the consistency of retrieval techniques on difficult queries~\cite{voorhees:2004}. It offers a traditional TREC-style evaluation setup of a document collection (528,155~documents from the Financial Times, the Federal Register~94, the LA Times, and FBIS), a set of topics (250~topics with title and description representing information needs and a narrative detailing which documents are considered relevant), and manual relevance judgments from human assessors. We use the short keyword-style titles (e.g., ``most dangerous vehicles'') as queries in our experiments and randomly sample 150~topics for training the neural rankers and the remaining~100 for our explainability experiments.

The MS MARCO dataset---originally only a collection for passage ranking and question answering---was recently used in the TREC~2019 and~2020 Deep Learning tracks since it provides large amounts of labeled data~\cite{mitra:2019}. We use the data from the document retrieval task that provides an end-to-end retrieval scenario for full documents similar to Robust04. The dataset consists of 3.2~million web documents and 367.000~training queries. Relevance labels are derived from the labels of the associated passage retrieval dataset under the assumption that a document with a relevant passage is a relevant document. While being a potential source of label noise, this does not constitute a hindrance to our experimental study, since we are not interested in evaluating the effectiveness of the trained models, but in explaining how they reach their ranking decisions. We index the concatenation of the URL, the title, and the body text of each document and randomly sample~100 of the~200 official test queries from the document retrieval task of the Deep Learning track for our explainability experiments---training of the neural ranking models described in the next section. Note that the MS~MARCO queries are notably longer (e.g., ``what are the effects of having low blood sugar'') than the Robust04 queries.

\subsection{Ranking Models}

In our study, we apply those neural ranking models on each dataset that have been prominently applied in the respective setting in previous work~\cite{guo:2016,pang:2016,mcdonald:2018,dai:2019,yilmaz:2019}---leaving a larger study with all rankers on both datasets for future work. For comparison, we also add three classical retrieval models.

On Robust04, we study three pairwise neural ranking models. \Ni MatchPyramid~\cite{pang:2016} uses a query-document interaction matrix as input to a convolutional neural network to extract matching patterns; we employ the cosine similarity variant (referred to as MP-COS henceforth), which performed best in preliminary experiments. \Nii DRMM~\cite{guo:2016} deploys a feed-forward neural network over query-document count histograms to output a relevance score. \Niii PACRR-DRMM~\cite{mcdonald:2018} combines the PACRR~\cite{hui:2017} model (which uses CNNs with different kernel sizes to extract position-aware signals from n-gram patterns) with the aggregation of~DRMM.

The neural models were set to re-rank the top-1000 results of~BM25. We trained DRMM and MP-COS using the MatchZoo toolkit~\cite{guo:2019}, fine-tuning according to the hyperparameters reported in the respective publications, and for PACRR-DRMM, we used its authors' implementation.\footnote{\url{https://github.com/nlpaueb/deep-relevance-ranking}} For all models, we used word embeddings fitted on Robust04, and labeled training data from the TREC Robust track. Our explainability experiments are run on the result sets of the 100~queries held out from the ranking model training.

On MS~MARCO, we study two BERT-based ranking models. \Ni DAI~\cite{dai:2019} is trained by fine tuning BERT to predict passage level relevance; we use the MaxP variant, where document relevance is the maximum of its passage-level relevances (DAI-MAXP). \Nii BERT-3S~\cite{yilmaz:2019} aggregates top-$k$ sentence scores, evaluated by a transfer model,\footnote{We fine-tuned this model using MS~MARCO; Microblogs were not used.} to compute document relevance scores.
Both models fine-tune a pretrained BERT model during training, and then use an aggregation procedure to predict document relevance; for more details on their training, we refer readers to the respective original publications. For our experiments, we used the MS~MARCO training set from the TREC~2019 Deep Learning track, and chose the best performing models according to~MAP calculated on the corresponding validation set. Our explainability experiments are run on the results sets of 100~queries from the MS~MARCO test set.

For the explainability experiments, we complement the neural ranking models with the classical retrieval models~BM25, TF-IDF, and~PL2---all parameters set to their defaults according to the implementation in the Anserini toolkit~\cite{yang:2017}.

%\begin{landscape}
%\centering
%\vspace*{30em}
\begin{table*}
\caption{Overview of the ranking explanation experiments. Explanation fidelity is measured as the proportion of document pairs ordered correctly, and is averaged over ten cross-validation folds across all models evaluated in the corresponding row.}
\label{table-experiments-overview}
%\small
\begin{tabular}{@{}lr@{\qquad}r@{\;}r@{\qquad}rrr@{\quad}rrr@{}}

\toprule
\multicolumn{2}{@{}l@{\qquad}}{\bfseries Explanation Models} & \multicolumn{2}{@{}l@{\qquad}}{\bfseries Instances per Model} & \multicolumn{6}{@{}c@{}}{\bfseries Explanation Fidelity} \\
\multicolumn{2}{@{}l@{\qquad}}{\bfseries Scope \hfill per Retr. Model} & \bfseries Train & \bfseries Test & \multicolumn{3}{@{}l@{\;}}{\bfseries Classical Retrieval Models} & \multicolumn{3}{@{}c@{}}{\bfseries Neural Retrieval Models} \\
%\midrule
%\multicolumn{9}{@{}l}{\emph{Global models}}\\

\midrule
\multicolumn{4}{@{}l}{\emph{Robust04}} & \bfseries BM25 & \bfseries TF-IDF & \bfseries PL2 & \bfseries MP-COS & \bfseries DRMM & \bfseries PACRR-DRMM\\
query                 & 100   & 8,943  & 1,279 & \bfseries 0.75 & \bfseries 0.66 & 0.78           & \bfseries 0.67 & \bfseries 0.68 & \bfseries 0.72  \\
rank-diff bin         & 24    & 38,213 & 4,380 & 0.71           & 0.63           & 0.77           & 0.59           & 0.61           & 0.67            \\
score-diff bin        & 24    & 38,327 & 4,265 & 0.72           & 0.64           & 0.78           & 0.59           & 0.61           & 0.68            \\
query, rank-diff bin  & 2,368 & 384    & 44    & 0.73           & 0.64           & 0.77           & 0.65           & 0.66           & 0.70            \\
query, score-diff bin & 2,394 & 383    & 44    & 0.74           & 0.65           & \bfseries 0.79 & 0.64           & 0.66           & 0.70            \\
\midrule
\multicolumn{4}{@{}l}{\emph{MS MARCO}} & \bfseries BM25 & \bfseries TF-IDF & \bfseries PL2 & & \bfseries BERT-3S & \bfseries DAI-MAXP\\
query                 & 100   & 8,936  & 1,278 & \bfseries 0.64 & \bfseries 0.60 & \bfseries 0.63 & & \bfseries 0.61 & \bfseries 0.59  \\
rank-diff bin         & 24    & 38,208 & 4,350 & 0.60           & 0.56           & 0.59           & & 0.57           & 0.54            \\
score-diff bin        & 24    & 38,280 & 4,278 & 0.61           & 0.56           & 0.59           & & 0.59           & 0.55            \\
query, rank-diff bin  & 2,400 & 382    & 44    & 0.62           & 0.58           & 0.61           & & 0.60           & 0.57            \\
query, score-diff bin & 2,376 & 386    & 44    & 0.63           & 0.60           & 0.62           & & 0.61           & 0.58            \\
\bottomrule
\end{tabular}
\end{table*}
%\end{landscape}

\subsection{Explanation Parameters}

For our experiments, we instantiate the axiomatic explanation framework described in Section~\ref{sec:approach} with 20~axioms: the 16~axioms shown in Table~\ref{table:axioms} and the aforementioned 4~embedding-based variants for the semantic-similarity axioms STMC1 and STMC2. We operate on the top-1000 results of each ranking, but sample only a subset of all constituent document pairs. Based on the assumption that explanations of the top ranks are of the most interest, we follow a non-uniform sampling strategy that includes all pairs of documents from the top-20 ranks, plus~1\%~of all remaining pairs sampled uniformly at random. We instantiate the explanation model as a random forest with 128~trees of maximum depth~20.

The input to the explanation model is a set of pairwise ranking decisions made by the retrieval model to be explained, at a scope that depends on the granularity of the explanation model (Section~\ref{sec:explanation-granularity}). This input dataset is randomly split into ten folds which are alternately used to fit the explanation model, and to evaluate its explanation fidelity, in a standard ten-fold cross-validation setting. Every individual instance comprises the identifiers of the two documents involved, the ranking preferences for this pair for each of the 20~axioms, and the ranking preference of the retrieval model to be explained. All experiments use the latter ranking preference as the dependent variable for the explanation model to predict and the axiomatic ranking preferences as independent variables.
Note that if an instance for document pair $(d_i, d_j)$ is included, so is the instance for pair~$(d_j, d_i)$---with inverted preferences. Both instances forming such mirrored pairs are always assigned to the same cross-validation fold to avoid train-test information leakage.

\subsection{Explanation Model Locality}
\label{sec:explanation-granularity}

To answer our research questions related to the degree of locality at which axiomatic explanations apply, we train explanation models not only at the scope of the full ranking, but also at the scope of subsets of the ranking. Overall, we consider three types of locality: \Ni locality by query, \Nii locality by ranking position, and \Niii locality by score differences. The distinction between ranking and score difference is useful to incorporate a notion of degree of certainty of the ranking model---a ranking assigns documents to different positions even if they obtain the same score. We create 24~bins of approximately the same size for locality by ranking position and for locality by score differences, and we combine the these with locality by query to obtain five binning strategies in total that we use to select documents to train our explanation models. 

The left-hand side of Table~\ref{table-experiments-overview} illustrates five different experiment configurations resulting from this setup, which fit explanation models at the following scopes: \Ni one model per query, yielding 100~explanation models per retrieval model and dataset; \Nii one model per one of the 24~bins of the ranking differences across all queries; \Niii one model per one of the 24~bins of min-max normalized differences in retrieval score across all queries; \Niv one model per combination of query and rank-difference bin; \Nv one model per combination of query and score-difference bin. The first two columns of Table~\ref{table-experiments-overview}~(``Scope'' and ``Per Retrieval Model'') show the scopes of the explanation models as described above, along with the number of explanation models per retrieval model resulting from the respective granularity level. The next two columns~(``Train'' and ``Test'') show the average number of training and test instances per cross-validation fold for each of these explanation models. Note that differences---mostly in the number of the most finely-granular explanation models---arise from the fact that not every retrieval model returns the full top-1000 results for every query. The values shown in the table are averaged over all retrieval models.
\section{Results}
\label{sec:results}

The ``Explanation Fidelity'' columns of Table~\ref{table-experiments-overview} show to what extent our axiomatic explanation framework can characterize three classical retrieval models (BM25, TF-IDF, and PL2) and several neural rankers on the Robust04 and MS~MARCO datasets. Explanation fidelity is measured as the accuracy of the explanation model in terms of reproducing the retrieval model's ranking decisions, macro-averaged over the ten cross-validation folds and over the number of explanation models (column ``per Retrieval Model'' in Table~\ref{table-experiments-overview}).

\subsection{Overall Explanation Fidelity}

We achieve explanation fidelities of~0.54 and higher for all examined ranking models under all considered parameters. The classical retrieval model~PL2 achieves the best explainability of nearly~80\% on the Robust04 dataset, where the explanation granularity makes little difference. This high accuracy indicates that the limited set of simple retrieval axioms---although individually comprehensible for humans---can be combined to explain at least some of these relatively complex ranking formulas. The BM25~model is nearly as accurately explainable as PL2, even though BM25 can be considered more complicated since it has two tunable parameters, and PL2 has none. Rankings produced with TF-IDF illustrate that the simplicity of the ranking model does not guarantee good explainability. The accuracy of the TF-IDF explanations varies only slightly over different explanation model scopes, which is an observation that repeats for all classical ranking models under consideration.

By contrast, the accuracy often varies more for different explanation scopes in case of rankings produced by neural models. We find that this does not pose a problem, since the simple query-level granularity is always a reasonable choice that often outperforms more complicated setups like binning by rank difference. In the end, the scope of the explanation model does not strongly influence the fidelity of the explanations for any retrieval model. Fitting one explanation model per ranking appears to be the most straight-forward approach, and in most cases the best-performing one.

The explanation fidelity of all neural ranking models under investigation on Robust04 is within the range of the fidelities obtained for the classical models TF-IDF and BM25. PACRR-DRMM reaches an accuracy of~0.72, almost at the level of BM25. Similarly, rankings from MP-COS and DRMM obtain accuracies slightly above TF-IDF, but both almost double the accuracy difference across model scopes.

On MS~MARCO, the explainability for all classical retrieval models is much lower than on Robust04, especially for~PL2. As outlined in Section~\ref{sec:collections}, the two collections significantly vary in terms of size and query characteristics, which may explain the drop in fidelity. The neural rankers tested on MS~MARCO also achieve poorer explanation fidelities compared to those tested on Robust04, but are not directly comparable. The BERT-3S model attains slightly better explanations than DAI-MAXP across all explanation model scopes. For all retrieval models tested on MS~MARCO, the query-scope explanation models perform best.

Due to our sampling strategy, the explanation model solves a balanced binary classification problem, where an accuracy of~0.5 corresponds to failure to explain the given ranking decisions better than random chance. As a sanity check, we apply the explanation model also to randomly-shuffled variants of the rankings from the previous experiment. In this setting, the explanation fidelity remains consistently below~0.52 everywhere.  Thus, our axiomatic framework explains all retrieval models at least somewhat better than random chance. However, especially for the MS~MARCO rankings, explanation fidelity is very limited. Going forward, we investigate the aggregated results from Table~\ref{table-experiments-overview} in finer detail to better understand in what contexts the axiomatic explanation models perform well, and which axioms are responsible.

\subsection{Explanation Fidelity by Score Difference}

\begin{figure*}[tb]
\centering%
\includegraphics[width=\columnwidth]{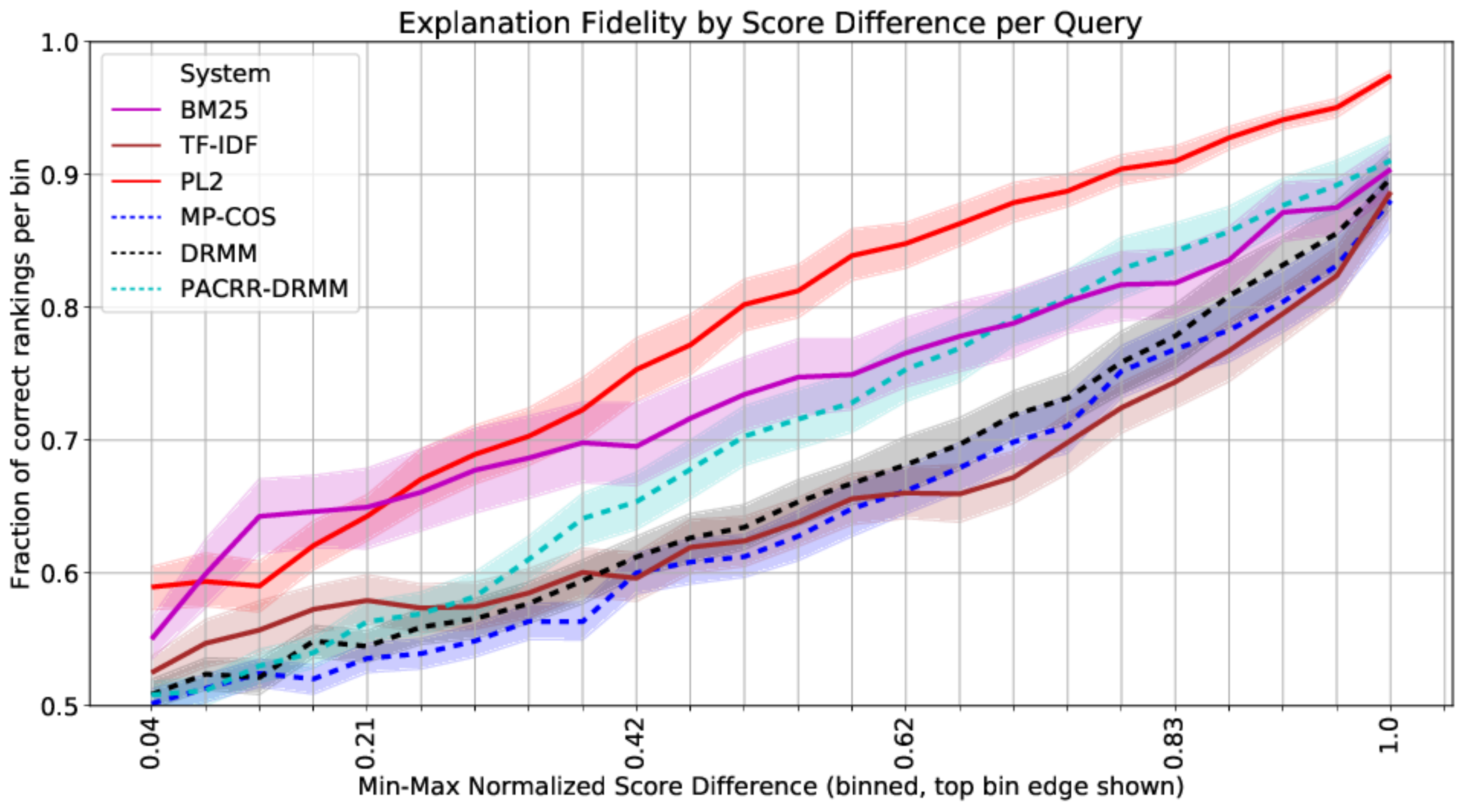}
%\hfill
\includegraphics[width=\columnwidth]{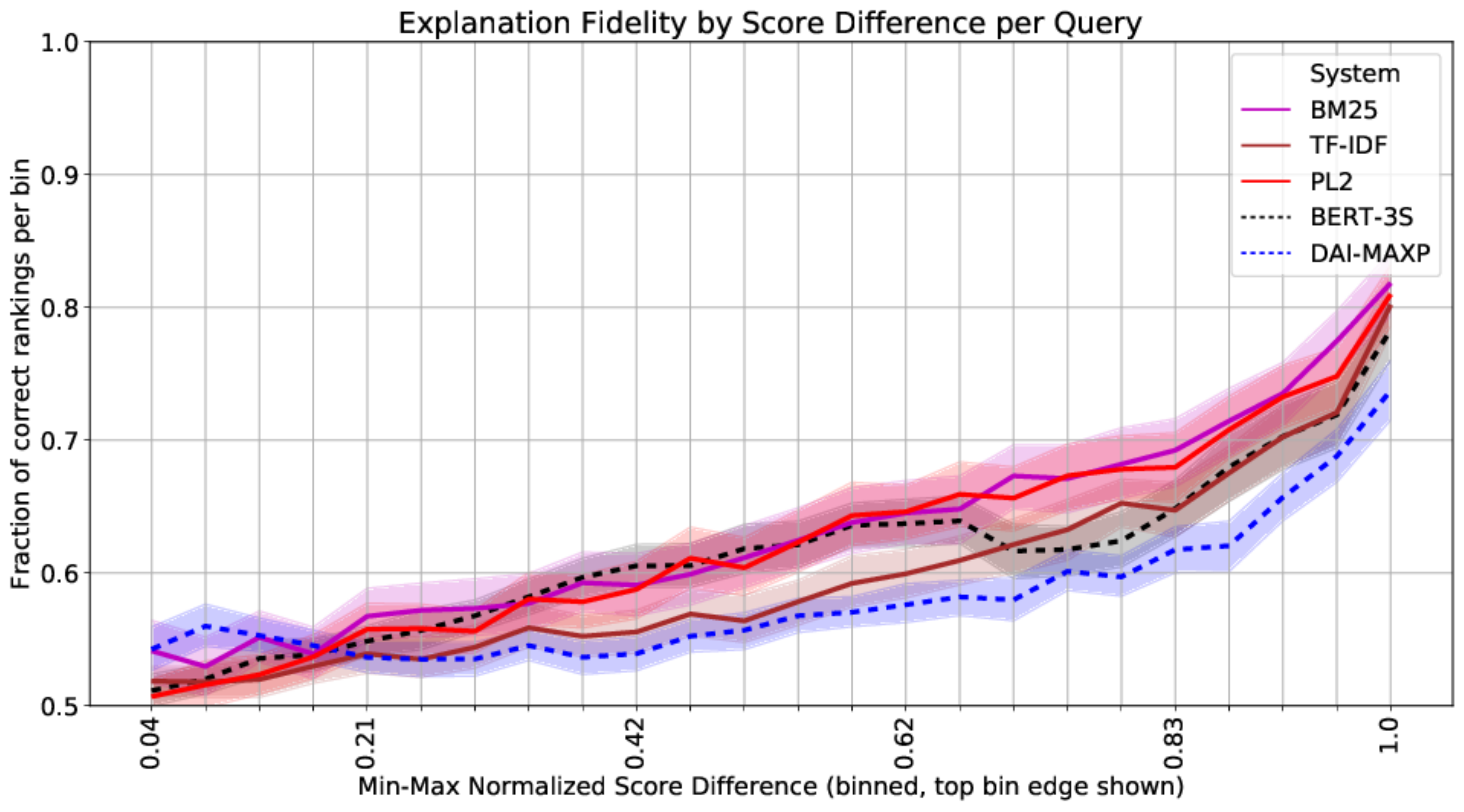}
\caption{Explanation fidelity by score difference per query on Robust04 (left) and MS~MARCO (right).}
\label{figure-score-diff-explainability-combined}%
\end{figure*}

An initial comparison of the explanation models at rank-difference scope to those at score-difference scope indicates that the score difference is slightly more useful both when training explanation models as well as when applying those models to explain rankings. This is expected, as a difference in rank does not necessarily correspond to a high confidence of the ranking model in the difference, since it does not take score ties into account~\cite{lin:2019}. For this reason, we proceed with the score-difference based binning.

To better understand how locality within the ranking might affect the fidelity of our axiomatic explanations, we inspect the performance of the query-level explanation models on the level of individual bins of the pairwise score differences produced by the studied retrieval models. For each dataset and retrieval model, we employ the~100 explanation models of the ``query'' scope (cf.\ Table~\ref{table-experiments-overview}) and subdivide the document pair instances from the test folds into~24 bins according to the min-max normalized difference in retrieval score. The bin edges are chosen in such a way that all bins contain the same number of instances on average across queries. We then compute the explanation fidelity for each bin separately and macro-average the results across the testing folds and queries.

Figure~\ref{figure-score-diff-explainability-combined} shows the explanation fidelity by the score difference binning per query as one line for each retrieval model's averaged explanation fidelities. Solid lines represent the classical retrieval models, and dashed lines the neural ones. The bands indicate 95\%~confidence intervals across queries. Small changes in a retrieval model's scores are hardly ever well explainable. This indicates that if the retrieval model is uncertain about a document pair and assigns a small score difference, this pair will also be hard to explain; large score differences, by contrast, will be much more explainable.

The increase in explainability with increasing score difference is more pronounced for Robust04; the explanations on both datasets perform pretty poorly at the low score difference end. This indicates that the better explainability of Robust04 rankings (Table~\ref{table-experiments-overview}) mainly results from document pairs far apart in retrieval score, while for the larger MS~MARCO collection the probably much larger clusters of highly-similarly scored documents may be more problematic.

For a deeper look into what the axiomatic preferences actually explain, we examine the feature importance of the axioms in the random forest explanation model. To this end, we fit the explanation models on four more coarse-grained bins over the min-max normalized retrieval scores, again chosen to contain approximately the same number of instances. In this setting, we examine which axioms account for small score differences of~12.6\% or less, middling differences from~12.6\% to~25.2\%, and from~25.6\% to~50\%, as well as large differences greater than~50\%. We perform this analysis on the Robust04 dataset due to its better overall explainability. To quantify the feature importance of an axiom~$A$, we measure the mean decrease in Gini impurity averaged over all decision trees in the ensemble, and over all splits where~$A$ is used, weighted by the number of training instances involved in the split~\cite{louppe:2013}.

Table~\ref{table-feature-importance} shows the mean decrease in impurity of the three most important axioms for each combination of score difference bin and retrieval model. 
Most strikingly, we found a large degree of overlap among nearly all investigated retrieval models, and across the majority of the score difference bins. The query aspect axiom REG features prominently; while this axiom rewards occurrence of the most ``outlier'' query term in result documents, the mere fact that it directly rewards query term occurrence may suffice to make it feature prominently in axiomatic explanations. The term proximity axiom PROX4 contributes to the axiomatic explanations in almost all cases. As originally formulated~\cite{stein:2016n}, PROX4 rewards the occurrence of close groupings of all query terms with few non-query terms in between. None of the classical retrieval models incorporate notions of term proximity, but meeting the PROX4 constraint implies that a document contains all query terms. The diversity axiom DIV, which penalizes Jaccard similarity to the query, constitutes a useful feature for most retrieval models and score differences, but since the explanation models are based on decision trees, they will likely use violations of the DIV axiom as a positive signal, as this is more in line with the behavior of the classical retrieval models. By further simplifying the explanation model to, for example, a regression model, such effects will be apparent in the algebraic signs of the parameters.

The rankings produced by the PL2 retrieval model, which tend to be best explainable overall on Robust04, tend to match the same axioms as other retrieval models for small score differences.
However, for large score differences of 50\% and more, PL2 has no overlap among the top three axioms with any other model, while all other models share the same top three axioms in this range. Interestingly, two variants of the semantic matching axiom STMC1 feature prominently for PL2 in this range, even though the PL2 scoring function considers exact matches only. It seems that the document property rewarded by STMC1---the presence of terms semantically similar to query terms---correlates with the divergence from randomness of the term frequency of query terms, which PL2 measures.

\subsection{Ranking Quality and Explanation Fidelity}

Over all Robust04 queries, we observe weak but consistently positive correlations between retrieval model effectiveness as measured by nDCG and the explanation fidelity of the query-scoped explanation models (Spearman's correlations of~0.16 for DRMM, 0.26~for MP-COS, and~0.2 for PACRR-DRMM). Table~\ref{table-robust04-fidelity-ndcg} shows some examples: the good rankings for queries like~425 or~612 tend to be much better explainable than the worse rankings for queries like~344 that are not well-explainable. However, there are notable exceptions like query~367 on which no system performs particularly well in terms of~nDCG but the explanation fidelity is high across all models.

\begin{table}[tb]
\centering
\footnotesize
\caption{Axiom importance for explanation fidelity on ROBUST04 in terms of mean decrease in impurity for the top three axioms across min-max normalized score difference bins (PACRR-DRMM denoted as PACRR for brevity).}
\label{table-feature-importance}
\begin{tabular}{@{}lr@{\hskip 0.045in}r@{\hskip 0.045in}r@{\hskip 0.045in}r@{\hskip 0.045in}r@{\hskip 0.05in}r@{\hskip 0.045in}r@{\hskip 0.045in}r@{\hskip 0.045in}r@{\hskip 0.05in}r@{\hskip 0.045in}r@{}}
\toprule
&\multicolumn{5}{c}{\bfseries Diff. $\leq 12.6\%$ } &\multicolumn{4}{c}{\bfseries   \hspace*{-0.35cm} $12.6\% < $ Diff. $\leq 25.2\%$} & &  \\
\cmidrule(r@{1.5em}){2-6}
\cmidrule(r@{1em}){7-10}

 & {\tiny\bfseries REG}  & {\tiny\bfseries PROX4}  & {\tiny\bfseries LNC1}  & {\tiny\bfseries DIV}  & {\tiny\bfseries LB1}  & {\tiny\bfseries REG}  & {\tiny\bfseries PROX4}  & {\tiny\bfseries LNC1}  & {\tiny\bfseries DIV}  &  \\
\midrule  

\bfseries \footnotesize BM25 & \footnotesize 0.14 & \footnotesize 0.13 & \footnotesize 0.17 &  &  & \footnotesize 0.16 & \footnotesize 0.16 & \footnotesize 0.13 &  & & \\
\bfseries \footnotesize TF-IDF & \footnotesize 0.15 & \footnotesize 0.12 &  &  & \footnotesize 0.12 & \footnotesize 0.15 & \footnotesize 0.12 & \footnotesize 0.15 &  &  & \\
\bfseries \footnotesize PL2 & \footnotesize 0.15 & \footnotesize 0.13 &  & \footnotesize 0.11 &  & \footnotesize 0.14 &  & \footnotesize 0.15 & \footnotesize 0.13 & & \\
\bfseries \footnotesize MP-COS & \footnotesize 0.16 & \footnotesize 0.11 &  & \footnotesize 0.11 &  & \footnotesize 0.17 & \footnotesize 0.11 &  & \footnotesize 0.11 &  & \\
\bfseries \footnotesize DRMM & \footnotesize 0.16 & \footnotesize 0.11 &  & \footnotesize 0.11 &  & \footnotesize 0.16 & \footnotesize 0.11 &  & \footnotesize 0.11 &  & \\
\bfseries \footnotesize PACRR & \footnotesize 0.14 & \footnotesize 0.12 &  & \footnotesize 0.12 &  & \footnotesize 0.15 & \footnotesize 0.14 &  & \footnotesize 0.11 &  & \\

\midrule
&\multicolumn{5}{c}{\bfseries $25.2\% < $ Diff. $\leq 50\%$} &\multicolumn{6}{c}{\bfseries $50\% < $ Diff. $\leq 100\%$} \\
\cmidrule(r@{1em}){2-6}
\cmidrule(r@{0.1em}){7-12}

&  {\tiny\bfseries REG}  & {\tiny\bfseries PROX4}  & {\tiny\bfseries PROX5}  & {\tiny\bfseries LNC1}  & {\tiny\bfseries DIV}  & {\tiny\bfseries REG}  & {\tiny\bfseries PROX4}  & {\tiny\bfseries PROX5}  & {\tiny\bfseries DIV}  & {\tiny\bfseries STMC1-f}  & {\tiny\bfseries STMC1-fr}\\
\midrule  

\bfseries \footnotesize BM25 & \footnotesize 0.21 & \footnotesize 0.16 & \footnotesize 0.09 &  &  & \footnotesize 0.28 & \footnotesize 0.16 & \footnotesize 0.15 &  &  & \\
\bfseries \footnotesize TF-IDF & \footnotesize 0.16 & \footnotesize 0.13 &  & \footnotesize 0.11 &  & \footnotesize 0.23 & \footnotesize 0.14 & \footnotesize 0.15 &  &  & \\
\bfseries \footnotesize PL2 & \footnotesize 0.11 &  &  & \footnotesize 0.11 & \footnotesize 0.20 &  &  &  & \footnotesize 0.25 & \footnotesize 0.13 & \footnotesize 0.13 \\
\bfseries \footnotesize MP-COS & \footnotesize 0.20 & \footnotesize 0.13 &  &  & \footnotesize 0.11 & \footnotesize 0.31 & \footnotesize 0.14 & \footnotesize 0.11 &  &  & \\
\bfseries \footnotesize DRMM & \footnotesize 0.19 & \footnotesize 0.12 & \footnotesize 0.11 &  &  & \footnotesize 0.28 & \footnotesize 0.14 & \footnotesize 0.18 &  &  & \\
\bfseries \footnotesize PACRR &  \footnotesize 0.20 & \footnotesize 0.15 & \footnotesize 0.11 &  &  & \footnotesize 0.29 & \footnotesize 0.15 & \footnotesize 0.17 &  &  & \\

\bottomrule
\end{tabular}
\vspace*{-0.5ex}
\end{table}

The top half of Table~\ref{table-ranking-examples} shows the first ten results of the DRMM ranking for query~367, along with our axiomatic explanations illustrated as a summary aggregating how many axioms would rank a particular result document in the same position, higher, or lower compared to the system ranking. Note that no axiom expresses a preference for every document pair---in fact, the total number of preferences is usually way lower than the actual number of 20~axioms. The last column illustrates the ranking preferences of individual axioms, as they relate to explaining the position of the document in the current row. Out of all axioms with any ranking preference regarding the row's document, we show the top three by how much they support or oppose the document's rank.

The many non-relevant top-10 results for query~367 can be explained by the specifics of the TREC topic. While DRMM only saw the single-term query ``piracy,'' the assessor instructions clarify that traditional high-seas piracy is meant, but not computer piracy as covered in many of the top-10 documents. The DRMM ranking is best explained by the REG and DIV axioms that may not really be the most ``intuitive'' choices for single-term queries. Still, REG degenerates to simply preferring documents with more query term occurrences for single-term queries which then ``makes sense'' for query~367. However, the agreement between the axioms and the top-10 ranks is rather poor; the overall good explanation fidelity comes from more distant pairs beyond the top~ten results. 

The lower half of Table~\ref{table-ranking-examples} shows the top-10 DRMM~results for the two-term query~425 ``counterfeiting money''. Here, DRMM performs considerably better and there also is a better agreement between the axioms and the top-10 ranks---with the term proximity axioms being the most important to explain the ranker's decisions.

%\begin{landscape}

\begin{table}[tb]
\caption{Explanation fidelity and retrieval effectiveness for selected Robust04 queries (PACRR-DRMM as PACRR).}
\label{table-robust04-fidelity-ndcg}
\footnotesize

\begin{tabular}{@{}l@{~}c@{~}c@{~}c@{}l@{\hspace{-5pt}}r@{}r@{}r@{}}
\toprule
\rotatebox[origin=l]{90}{\hspace{-1em}\bfseries Query} & \multicolumn{3}{@{}l@{}}{\bfseries Explanation fidelity} & \bfseries Topic title                & \multicolumn{3}{@{}c@{}}{\bfseries nDCG} \\[0.5ex]
 & MP-COS   & DRMM~ & PACRR & & MP-COS~ & DRMM~ & PACRR \\
\midrule
344   &   0.55 &     0.57 &       0.60 &                 Abuses of E-Mail &   0.27 & 0.26 &       0.33 \\
352   &   0.56 &     0.64 &       0.61 &           British Chunnel impact &   0.17 & 0.15 &       0.15 \\
356   &   0.73 &     0.61 &       0.60 &   Postmenopaus. estrogen Britain &   0.10 & 0.15 &       0.10 \\
367   &   0.74 &     0.82 &       0.87 &                           Piracy &   0.34 & 0.29 &       0.31 \\
399   &   0.61 &     0.57 &       0.71 &            Oceanographic vessels &   0.28 & 0.30 &       0.28 \\
409   &   0.84 &     0.81 &       0.62 &               Legal, Pan Am, 103 &   0.24 & 0.35 &       0.44 \\
425   &   0.78 &     0.81 &       0.85 &             Counterfeiting money &   0.74 & 0.77 &       0.75 \\
612   &   0.80 &     0.81 &       0.85 &                 Tibet protesters &   0.46 & 0.57 &       0.54 \\
618   &   0.81 &     0.82 &       0.83 &         Ayatollah Khomeini death &   0.52 & 0.41 &       0.44 \\
684   &   0.55 &     0.67 &       0.66 &               Part-time benefits &   0.26 & 0.41 &       0.39 \\
\bottomrule
\end{tabular}

\end{table}

%\end{landscape}

%\begin{landscape}
%\centering
%\vspace*{30em}
\begin{table*}
\caption{Example rankings with axiomatic explanations. (Non-)relevant Document IDs have a suffix~$+$~($-$). Axiomatic explanations show how many axioms would rank this result the same ($\Leftrightarrow$), higher ($\uparrow$) or lower ($\downarrow$), followed by up to three most relevant axioms, and how many other results they rank the same (superscript) or differently (subscript).
}
\label{table-ranking-examples}
\small
\providecommand{\AXSP}[3]{\ensuremath{%
\mathrm{#1}^{\textcolor{green!40!black}{#2}}_{\textcolor{red!75!black}{#3}}%
}}
\begin{tabular}{@{}rl@{~}lll@{}}
\toprule
    \multicolumn{3}{@{}l@{}}{
    {\bfseries System:}~DRMM
    \qquad
    {\bfseries Query:}~367
    \quad
    ``piracy''
    }
    &
   \multicolumn{2}{l@{}}{\bfseries Axiomatic explanation} 
    \\[.5ex]
\multicolumn{2}{@{}l}{\bfseries Rank Docid}&
\bfseries Content &
\bfseries \tiny $ \Leftrightarrow/~\uparrow/~\downarrow $ &
\bfseries Most relevant axioms \\
\midrule
 1 &  FT944-16684$^{-}$ &  Software companies offer rewards in anti-piracy drive -- Leading software compan\dots &  1 / 0 / 1 &  $\AXSP{REG}{9}{0}$\quad $\AXSP{DIV}{3}{6}$ \\
 2 &  FT931-8281$^{-}$ &  Survey of Personal and Portable Computers (21): Tougher times for pirates / a lo\dots &  2 / 0 / 1 &  $\AXSP{REG}{8}{1}$\quad $\AXSP{DIV}{4}{5}$\quad $\AXSP{LNC1}{1}{0}$ \\
 3 &  FT934-14966$^{-}$ &  Sixties buccaneer deals in high-tech piracy -- In his days as a radio caroline p\dots &  3 / 0 / 3 &  $\AXSP{DIV}{4}{5}$\quad $\AXSP{REG}{3}{5}$\quad $\AXSP{STMC2}{1}{0}$ \\
 4 &  FT911-1567$^{-}$ &  A rich haul from the sound of music: the illicit copying and sale of recorded mu\dots &  3 / 1 / 0 &  $\AXSP{REG}{7}{2}$\quad $\AXSP{DIV}{4}{5}$\quad $\AXSP{LNC1}{1}{0}$ \\
 5 &  LA032889-0045$^{-}$ &  Busting cable pirates; simi valley piracy case is one of the first to result in \dots &  3 / 1 / 0 &  $\AXSP{REG}{6}{1}$\quad $\AXSP{DIV}{4}{5}$\quad $\AXSP{LNC1}{2}{0}$ \\
 6 &  FBIS3-43017$^{-}$ &  Computer piracy in russia is a widespread phenomenon. the world average ratio of\dots &  5 / 1 / 0 &  $\AXSP{REG}{6}{1}$\quad $\AXSP{DIV}{3}{5}$\quad $\AXSP{LNC1}{1}{0}$ \\
 7 &  FBIS3-42979$^{-}$ &  Computer piracy in russia is a widespread phenomenon. the world average ratio of\dots &  2 / 1 / 0 &  $\AXSP{REG}{6}{1}$\quad $\AXSP{DIV}{3}{5}$\quad $\AXSP{LNC1}{1}{0}$ \\
 8 &  FT923-9880$^{+}$ &  Jakarta sinks plan to combat piracy -- Plans for an international centre to figh\dots &  5 / 0 / 0 &  $\AXSP{REG}{8}{1}$\quad $\AXSP{DIV}{8}{1}$\quad $\AXSP{STMC2}{2}{0}$ \\
 9 &  FT944-9277$^{-}$ &  UK company news: BSkyB says piracy could undermine float confidence -- British s\dots &  2 / 1 / 0 &  $\AXSP{DIV}{8}{1}$\quad $\AXSP{REG}{8}{0}$\quad $\AXSP{LNC1}{0}{1}$ \\
 10 &  FT924-15875$^{-}$ &  Piracy warnings -- A 24-hour centre to counter piracy in the seas of south-east \dots &  2 / 0 / 0 &  $\AXSP{DIV}{9}{0}$\quad $\AXSP{REG}{9}{0}$
  \\[1ex]
%%%%%%%%%%%%%%%%%%%%%%%%%%%%%%%%%%%%%%%%%%%%%%%%%%%%%%%%%%%%%%%%%%%%%%%%
% ranking 2
%%%%%%%%%%%%%%%%%%%%%%%%%%%%%%%%%%%%%%%%%%%%%%%%%%%%%%%%%%%%%%%%%%%%%%%%
\toprule
    \multicolumn{3}{@{}l@{}}{
    {\bfseries System:}~DRMM
    \qquad
    {\bfseries Query:}~425
    \quad
    ``counterfeiting money''
    } &
\bfseries \tiny $ \Leftrightarrow/~\uparrow/~\downarrow $ &
\bfseries Most relevant axioms \\
\midrule
 1 &  FBIS3-58171$^{+}$ &  The head of the gang that wanted to circulate some 970,000 counterfeit dollars i\dots &  6 / 0 / 5 &  $\AXSP{PROX5}{8}{1}$\quad $\AXSP{PROX3}{7}{2}$\quad $\AXSP{STMC1\_f}{7}{2}$ \\
 2 &  FBIS4-46741$^{+}$ &  Crime Counterfeiting is probably one of the world's oldest and most widespread t\dots &  10 / 0 / 1 &  $\AXSP{PROX2}{7}{2}$\quad $\AXSP{PROX3}{7}{2}$\quad $\AXSP{REG}{7}{2}$ \\
 3 &  FBIS4-26260$^{+}$ &  The Hongqiao District people's court examined and concluded a case of traffickin\dots &  5 / 0 / 5 &  $\AXSP{PROX4}{7}{2}$\quad $\AXSP{PROX5}{7}{2}$\quad $\AXSP{PROX2}{7}{2}$ \\
 4 &  LA091590-0091$^{+}$ &  PLUMBERS DISCOVER CASH FLOW PROBLEM IN SEWER; COUNTERFEITING:\dots &  7 / 2 / 0 &  $\AXSP{PROX4}{7}{2}$\quad $\AXSP{PROX5}{7}{2}$\quad $\AXSP{PROX1}{6}{3}$ \\
 5 &  FBIS3-54773$^{+}$ &  On 25 December a criminal gang of six Chechnya inhabitants was arrested in St. P\dots &  7 / 1 / 2 &  $\AXSP{PROX4}{7}{2}$\quad $\AXSP{PROX5}{7}{2}$\quad $\AXSP{REG}{5}{4}$ \\
 6 &  LA102189-0077$^{+}$ &  CALIFORNIA IN BRIEF; MODESTO; MAN INDICTED IN COUNTERFEITING\dots &  7 / 3 / 0 &  $\AXSP{PROX4}{7}{1}$\quad $\AXSP{PROX5}{6}{3}$\quad $\AXSP{PROX1}{6}{3}$ \\
 7 &  FBIS4-47199$^{+}$ &  The number of counterfeit ruble bank notes, bank notes of convertible currency, \dots &  7 / 4 / 0 &  $\AXSP{PROX5}{8}{1}$\quad $\AXSP{STMC1}{6}{3}$\quad $\AXSP{PROX4}{6}{1}$ \\
 8 &  FBIS4-58263$^{+}$ &  Counterfeit, 1990-issue \$100 bills have recently found their way onto the Jorda\dots &  6 / 4 / 0 &  $\AXSP{PROX4}{7}{2}$\quad $\AXSP{REG}{7}{2}$\quad $\AXSP{PROX3}{7}{1}$ \\
 9 &  FBIS4-59139$^{+}$ &  For some time, Tel Aviv has anxiously been raising with Egyptian politicians an \dots &  10 / 1 / 0 &  $\AXSP{PROX1}{8}{1}$\quad $\AXSP{PROX3}{8}{1}$\quad $\AXSP{PROX2}{8}{1}$ \\
 10 &  LA010390-0055$^{-}$ &  RIVAL ATHLETIC SHOE MAKERS FORMED AN ALLIANCE TO BATTLE COUNT\dots &  6 / 5 / 0 &  $\AXSP{PROX1}{9}{0}$\quad $\AXSP{PROX4}{8}{1}$\quad $\AXSP{PROX5}{8}{1}$ \\
\bottomrule

\end{tabular}
\end{table*}
%\end{landscape}

\subsection{Limitations and Summary}

The results of the preceding analysis indicate that while good explanation fidelity is possible in some contexts, the reason why axiomatic explanations work at all can be somewhat incidental to the criteria employed by the retrieval model to be explained. In this sense, not much can yet be said on what brings about the ranking decisions of the more inscrutable neural rankers, except that similar criteria to classical retrieval scenarios clearly apply. However, we also do find evidence that something else is at play. In an experiment with versions of the ORIG axiom~\cite{stein:2016n} expressing the ranking preferences of the classical retrieval models, all the classical models are highly effective at explaining each other's decisions. The full axiom set of the 20~before mentioned axioms extended by two axioms each expressing the rank preferences of the respective two not-to-be-explained classical models reaches explanation fidelities above~99\% for the classical models. Still, the explanation fidelity for the neural models increases only moderately.

We thus hypothesize that formulations we did use for the 20~axioms still lack some reliable indicators for some basic relevance signals used by classical retrieval models. For instance, even though there are axioms capturing term frequency, their current formulation might not be very useful in practice. To test this, we investigate how often the individual axioms' preconditions are satisfied, and find that the 10\%~relaxation of equality constraints---we used it following Hagen~et~al.~\cite{stein:2016n}---may still be too strict. In an experiment on Robust04, we find that the term frequency axioms' document length precondition is satisfied only in approximately~7\% of the document pairs, the preconditions of LNC1 only in~9\%, the proximity axioms PROX1--3 can be applied to only~21\%, and axiom LB1 to only 36\%~of the document pairs, while the remaining axioms apply to~90\% or more of the document pairs.

In summary, we find that large differences in retrieval score can be reasonably well explained with the simple axiomatic feature set employed in our study. Especially for the MS~MARCO dataset, the explainability does not depend very much on the specific retrieval model used to produce the ranking, and across all datasets, the simple setup of training one explanation model per query outperforms more complicated binning approaches, although binning may still be useful to understand model behavior across levels of score differences. However, a closer investigation of this behavior indicates that our current axiom set does not fully capture the scoring criteria of most ranking models, one possibly reason being the strict preconditions contained in several of the axioms.

\section{Conclusion \& Future Work}
\label{sec:conclusion}

We have introduced an axiomatic framework to explain the result rankings of information retrieval systems in terms of how well a system's ranking decisions adhere to a set of axiomatic constraints. Instantiated with a set of 20~axioms from the literature and a random forest model to reconstruct pairwise orderings from axiomatic ranking preferences, we have demonstrated our suggested framework's general capacity to explain rankings in an experimental study on the Robust04 and MS~MARCO test collections. The results show that axiomatic explanations for eight different retrieval systems---five of them complex deep neural network-based ranking functions and three classical scoring functions---work reliably for document pairs with very different retrieval scores (i.e., corresponding to a high confidence in a difference in relevance). Pairs with more similar retrieval scores are more difficult to explain---not too surprising given the rather few retrieval aspects that the 20~axioms do cover. Especially axioms with a precondition constraining the documents' length difference can rarely be applied, even when this constraint is relaxed to allow for a 10\%~difference, as was suggested in previous studies. Further relaxing or even dropping preconditions entirely may be an easy remedy, but also a vast departure from the original axioms and their formalization of the constraints they capture. Instead, it seems desirable to formulate new axioms that capture the same ideas in a more practically applicable way, or that capture retrieval constraints not yet covered by the known axioms, and to develop a weighting scheme that can quantify the degree to which preconditions are satisfied.

The explanation fidelity on the smaller, more genre-focused Robust04 collection with its shorter queries is superior to that on the MS~MARCO dataset. Further investigation into the causes of this discrepancy is warranted, but the vastly different characteristics of the respective queries and documents seem likely candidates. The explainability of neural rankers is mostly on par with that of classical retrieval functions, and there are notable overlaps in the axioms that are most useful to the explanation models. Still, the ``known'' and studied axioms do not cover a range of aspects important to modern search engines such as the timeliness of results, stylistic and readability considerations, or how well some results match user preferences expressed through previous interactions. 

While formalizing and operationalizing axiomatic constraints for such properties certainly seems worthwhile, such an endeavor was beyond the scope of our paper. Even though we can demonstrate promising first steps to axiomatically explain retrieval systems' result rankings, the addition of further well-grounded axiomatic constraints capturing other retrieval aspects seems to be needed to further improve the explanations. Its current limitations notwithstanding, we consider our approach a promising complement to the more tightly-controlled studies from previous work~\cite{rennings:2019,camara:2020,macavaney:2020}. While the latter shed light on the general principles under which complex relevance scoring models operate, our axiomatic reconstruction framework could help IR system designers---or even end users---make sense of a concrete ranking for a real-world query.

\begin{acks}
This work has been partially supported by the DFG through the project ``ACQuA: Answering Comparative Questions with Arguments'' (grant HA 5851/2-1) as part of the priority program ``RATIO: Robust Argumentation Machines'' (SPP 1999). Jaspreet Singh's contributions were made prior to his affiliation with Amazon.
\end{acks}

\bibliographystyle{ACM-Reference-Format}
%\bibliography{ictir21-axiomatic-explanations-lit}

\end{document}